\documentclass[12pt]{article}

\usepackage{graphicx}
\usepackage{amsmath, amssymb}
\usepackage{slashed}
  
\begin{document}

\begin{titlepage}

\def\thefootnote{\fnsymbol{footnote}}

\begin{center}

\hfill DESY 14-114
\hfill  \\
\hfill July, 2014

\vspace{0.5cm}
{\large \bf  Axino Dark Matter 
in Moduli-induced Baryogenesis}

\vspace{1cm}
Koji Ishiwata

\vspace{0.5cm}

{\it {Deutsches Elektronen-Synchrotron DESY, 
         22607 Hamburg, Germany}}

\vspace{1cm}

\abstract{We consider axino dark matter in large $R$-parity violation
  (RPV).  In moduli-dominated universe, axino is produced thermally or
  non-thermally via saxion decay, then late-decaying moduli dilute
  axino density, which results in the right abundance to explain the
  present dark matter. At the same time baryon asymmetry is generated
  due to moduli-induced baryogenesis via the large RPV.  Axino is
  cosmologically stable in spite of the large RPV since its decay rate
  is suppressed by the axion decay constant, heavy squark mass or
  kinematics.  }

 \end{center}
 \end{titlepage}

\renewcommand{\theequation}{\thesection.\arabic{equation}}
\renewcommand{\thepage}{\arabic{page}}
\setcounter{page}{1}
\renewcommand{\thefootnote}{\#\arabic{footnote}}
\setcounter{footnote}{0}
 
\section{Introduction}
\label{sec:intro}
\setcounter{equation}{0} 

The discovery of the Higgs boson at the LHC confirmed the standard
model of particle physics~\cite{Aad:2012tfa,Chatrchyan:2012ufa}.  So
far no phenomenon, which shows severe inconsistency with the standard
model (SM), has been reported on the ground-based experiments (except
for neutrino oscillation).  In cosmology, however, it is clear that we
need new physics beyond the standard model. First of all, the standard
cosmology can not explain the existence of dark matter (DM). In
addition, the baryon density predicted in the standard model is too
small to account for the observed value.  Supersymmetry (SUSY) is a
promising solution to the issues. On top of that, string theory, which
requires supersymmetry for the consistency, is a viable candidate for
the theory of everything.

However, such an extension may cause another problem especially in
cosmology. Moduli fields, which must be stabilized to compactify the
extra dimensions in string theory, may be destabilized during
inflation if the inflation scale is very high, which is indicated by
the recent BICEP2 observation~\cite{Ade:2014xna}. Even if the
destabilization is avoided in some
ways~\cite{Kallosh:2004yh,BlancoPillado:2005fn}, it is likely that
moduli are displaced far from their true minima at the end of
inflation.  Then moduli start to oscillate, and soon dominate the
energy density of the universe. The moduli-dominated universe ends
when moduli decay, accompanying a huge entropy injection.  This is
potentially problematic because such substantial an entropy production
dilutes pre-existing matter density, then it is difficult to lead to
big bang nucleosynthesis (BBN) and the structure formation of the
universe.  Possible way out are following: production of a large
amount of the matter density before moduli decay or generation of the
matter density after moduli decay. As for baryonic matter,
Affleck-Dine mechanism~\cite{Affleck:1984fy,Dine:1995kz} is a typical
example of the former one. On the other hand, late-decaying
gravitino~\cite{Cline:1990bw} or saxion~\cite{Mollerach:1991mu} can
also produce baryon asymmetry, which corresponds to the latter.
Recently another mechanism, moduli-induced baryogenesis, was
proposed~\cite{Ishiwata:2013waa}.  It was shown that
Kachru-Kallosh-Linde-Trivedi scenario~\cite{Kachru:2003aw} has
built-in features for baryogenesis, such as large enough CP phase and
suitable mass spectrum for superparticles.  Then subsequent decays of
gluino and squarks from moduli produce sufficient baryon asymmetry.
In those baryogenesis due to late-decaying particle, however, a large
$R$-parity violation (RPV) is assumed, which makes lightest
superparticle (LSP) unstable. This is a downside to accounting for
dark matter.

In this paper we consider axino LSP in large $R$-parity violation.
Introducing the axion supermultiplet is motivated by Peccei-Quinn (PQ)
mechanism~\cite{Peccei:1977hh}, which solves the strong CP problem.
Assuming that the fermionic component of the axion multiplet, axino,
is the LSP, axino is copiously produced by its radial component,
saxion, decay and the scattering from thermal plasma. Axino can be
cosmologically stable even if the RPV is $O(1)$ because its decay rate
is suppressed by the axion decay constant, squark mass or kinematics.
After saxion decay, moduli decay follows to dilute axino abundance,
which results in the observed relic of dark matter. At the same time,
baryon asymmetry of the universe is generated in moduli-induced
baryogenesis with the RPV.

\section{Cosmological scenario}
\label{sec:scenario}
\setcounter{equation}{0} 

In this section we describe the basic picture of our scenario.  In the
scenario moduli dominate the total energy of the universe after
inflation.  Axino is produced thermally or by non-thermal saxion decay
in the epoch of moduli domination. Eventually moduli decay and dilute
the axino abundance, which gives the right value to explain DM. Here
baryon asymmetry is generated from moduli decay as well due to
$R$-parity violated interaction.  The saxion decay also generates
axion. Although it is diluted by moduli decay, the produced axion may
give a sizable contribution to radiation as dark radiation.  Finally
the stability of axino under the RPV is discussed.

\subsection{Moduli-dominated universe}
\label{sec:XDuniverse}

Let us begin with the thermal history after inflation.  As we
mentioned in the Introduction, the modulus field tends to be displaced
from its true minimum due to the deformed potential during inflation
or due to the initial condition.  Then after inflation, it starts to
oscillate around the true minimum when the Hubble parameter $H$
reduces to moduli mass $m_X$.  Assuming $T_R$, the reheating
temperature after inflation, is comparable to $T_{X,{\rm osc}}$, the
temperature when modulus begins to oscillate, the energy density of
modulus field $X$ per entropy density freezes after the oscillation
starts at a value of
\begin{eqnarray}
\frac{\rho_X(T)}{s(T)}=
\frac{1}{8}\,T_{X,{\rm osc}}
\left(\frac{\delta X_{\rm ini}}{M_{\rm Pl}}\right)^2
\equiv
\left[\frac{\rho_X}{s}\right]_{\rm osc}.
\label{eq:rhoX/s_osc}
\end{eqnarray}
Here $T$ is the cosmic temperature, $\rho_X$ is the energy density of
modulus, $s(T)$ is the entropy density and $\delta X_{\rm ini}$ is the
initial amplitude of $X$ measured from its true minimum. Typically we
expect $\delta X_{\rm ini}\sim M_{\rm Pl}$ where $M_{\rm Pl}\simeq
2.4\times 10^{18}\,{\rm GeV}$ is the reduced Planck mass.  $T_{X,{\rm
    osc}}$ is estimated from the equation $H\simeq m_X$ as
\begin{eqnarray}
T_{X,{\rm osc}}&=& 
\left[\frac{90}{\pi^2g_*(T_{X,{\rm osc}})}\right]^{1/4}\sqrt{M_{\rm Pl}m_X}
\nonumber \\
&\simeq& 6.9\times 10^{13}\,{\rm GeV} 
\left(\frac{m_X}{10^{10}\,{\rm GeV}}\right)^{1/2}.
\label{eq:TXosc}
\end{eqnarray}
Here $g_*(T)$ counts degree of freedom of relativistic particles in
the thermal bath. Due to their huge energy density, moduli soon
dominate the energy density of the universe.  The temperature when
moduli begin to dominate the universe is estimated from the relation
$\rho_X \simeq \rho_R$ as
\begin{eqnarray}
T_{\rm dom}\simeq \frac{1}{6}T_{X,{\rm osc}}
\left(\frac{\delta X_{\rm ini}}{M_{\rm Pl}}\right)^2,
\label{eq:Tdom}
\end{eqnarray}
where we have used Eq.~(\ref{eq:rhoX/s_osc}).  It is seen moduli
dominate the total energy density soon after starting to oscillate.

Since the energy density of moduli redshift as $\rho_X \propto a^{-3}$,
it is given by
\begin{eqnarray}
\rho_X(T)
=\left[\frac{\rho_X}{s}\right]_{\rm osc}s(T),
\end{eqnarray}
until moduli decay. As $\rho_{\rm tot}$, the total energy density of
the universe, is equal to $\rho_X$ during moduli domination, the
Hubble parameter in moduli-dominated universe is given by
\begin{eqnarray}
H\simeq \sqrt{\frac{\rho_{\rm tot}}{3M_{\rm Pl}^2}}\simeq 
\frac{\sqrt{T_{\rm dom}s(T)}}{2M_{\rm Pl}}.
\label{eq:H_XD}
\end{eqnarray}

The epoch of moduli domination terminates when moduli decay to
particles in minimal supersymmetric standard model (MSSM), and it
turns into radiation domination.  The temperature at the beginning
of this radiation-dominated universe is determined by $H\simeq
\Gamma_X$ as
\begin{eqnarray}
T_X&=& 
\left[\frac{90}{\pi^2g_*(T_X)}\right]^{1/4}\sqrt{M_{\rm Pl}\Gamma_X}
\nonumber \\
&\simeq& 9.8 \times 10^{4}\,{\rm GeV}
\left(\frac{m_X}{10^{10}\,{\rm GeV}}\right)^{3/2}.
\label{eq:TX}
\end{eqnarray}
Here we have used the decay rate of moduli, which is given
by~\cite{Endo:2006zj}
\begin{eqnarray}
\Gamma_X\simeq \frac{c_X}{4\pi}\frac{m_X^3}{M_{\rm Pl}^2},
\label{eq:GammaX}
\end{eqnarray}
where $c_X$ is $O(1)$ constant and here and hereafter we take it as
unity.\footnote{In the numerical analysis, we use the results given in
  Ref.~\cite{Ishiwata:2013waa}.}  The moduli masses have to be larger
than around $100~{\rm TeV}$ in order not to destroy BBN. Even if $m_X
\gtrsim 100~{\rm TeV}$ is satisfied, however, a huge entropy
production due to moduli decay may strongly dilute primordial relics,
such as baryon and DM. The effect is described by a dilution factor,
which is given by a ratio of entropy density before and after the
moduli decay,
\begin{eqnarray}
d_X&=& \frac{3}{4}\,T_X
 \left[\frac{\rho_X}{s}\right]^{-1}_{\rm osc}
= 6\,\frac{T_{X}}{T_{X,{\rm osc}}} 
\left(\frac{M_{\rm Pl}}{\delta X_{\rm ini}}\right)^2
\nonumber \\
&\simeq& 8.5 \times 10^{-9} 
\left(\frac{m_X}{10^{10}\,{\rm GeV}}\right)
\left(\frac{M_{\rm Pl}}{\delta X_{\rm ini}}\right)^2.
\label{eq:dX}
\end{eqnarray}
The dilution is important to reduce over-produced axino (and axion),
which is discussed below.

\subsection{Saxion decay}

Saxion is the radial component field in the axion supermultiplet.  The
axion supermultiplet is determined as a flat direction of the scalar
potential given by the PQ fields. Here the PQ fields have non-zero PQ
charges and break PQ symmetry spontaneously.  We define the axion
supermultiplet as
\begin{eqnarray}
A=\frac{1}{\sqrt{2}}(\sigma+i \,a)+\sqrt{2}\theta \tilde{a}
+F{\mathchar`-{\rm term}}.
\end{eqnarray}
Here $\sigma$, $a$ and $\tilde{a}$ are saxion, axion and axino,
respectively. For later calculation we define the axion decay constant
as $f_a=\sqrt{2\sum_i q_i^2 v_i^2}$ where $q_i$ and
$v_i=\langle{\Phi}_i\rangle$ are the PQ charge and the vacuum
expectation value (VEV) of a PQ field $\Phi_i$, respectively. If the
domain wall number $N_{\rm DW}$ is not unity, then $f_a$ should be
$\sqrt{2\sum_i q_i^2 v_i^2}/N_{\rm DW}$.

Similar to moduli, saxion tends to have initial amplitude around its
true minimum after inflation then it starts oscillation when $H\simeq
m_\sigma$ ($m_\sigma$ is saxion mass).  Around this period, moduli
begin to dominate the total energy. If saxion starts to oscillate
before moduli domination, the temperature at the beginning of the
oscillation is given by
\begin{eqnarray}
T_{\sigma,{\rm osc}}&\simeq& 
\left[\frac{90}{\pi^2g_*(T_{\sigma,{\rm osc}})}\right]^{1/4}
\sqrt{M_{\rm Pl}m_\sigma},
\label{eq:Tsigmaosc}
\end{eqnarray}
and its energy density to entropy ratio is fixed at
\begin{eqnarray}
\left[\frac{\rho_\sigma}{s}\right]_{\rm osc}=
\frac{1}{8}T_{\sigma,{\rm osc}}
\left(\frac{\delta \sigma_{\rm ini}}{M_{\rm Pl}}\right)^2.
\label{eq:rho_sigma1}
\end{eqnarray}
Here $\delta \sigma_{\rm ini}$ is the saxion initial amplitude, which
is expected to be order of $f_a$ to $M_{\rm Pl}$. The value depends on
the saxion potential (see, {\it e.g.},
Refs.~\cite{Kawasaki:2011ym,Moroi:2012vu,Moroi:2013tea}).  On the
other hand, saxion starts to oscillate after the universe is dominated
by moduli when $T_{\sigma,{\rm osc}}<T_{\rm dom}$.  In such a case,
saxion energy density per entropy density has a fixed value
\begin{eqnarray}
\left[\frac{\rho_\sigma}{s}\right]_{\rm osc}=
\frac{1}{8}T_{\rm dom}
\left(\frac{\delta \sigma_{\rm ini}}{M_{\rm Pl}}\right)^2.
\label{eq:rho_sigma2}
\end{eqnarray}
In our scenario we consider $m_X$ is relatively larger than
$m_{\sigma}$. Then the energy density of moduli is much larger than
that of saxion during the period of moduli domination in either case.

After the coherent oscillation, saxion decays to lighter particles.
The decay rate depends on axion model, {\it i.e.},
Kim-Shifman-Vainshtein-Zakharov (KSVZ) (or hadronic axion)
model~\cite{Kim:1979if} or Dine-Fischler-Srednicki-Zhitnitsky (DFSZ)
model~\cite{Dine:1981rt}. In both KSVZ and DFSZ models saxion couples
to axino and axion via the kinetic term, which is given
as~\cite{Chun:1995hc}
\begin{eqnarray}
{\cal L}_{\sigma}=\left(1+\frac{2\xi}{f_a}\sigma\right)
\left[
\frac{1}{2} (\partial_\mu \sigma)^2
+ \frac{1}{2} (\partial_\mu a)^2
+\frac{1}{2}\bar{\tilde{a}}i \slashed{\partial}\tilde{a}
\right],
\end{eqnarray}
where $\xi=2\sum_i q_i^3v_i^2/f_a^2$. Here we have used $\tilde{a}$ as
a four component axino spinor.  From this interaction, the partial
decay widths for $\sigma \rightarrow aa$ and $\sigma \rightarrow
\tilde{a}\tilde{a}$ are computed as
\begin{eqnarray}
\Gamma(\sigma \rightarrow a a)
&=&\frac{\xi^2}{32\pi} \frac{m_\sigma^3}{f_a^2},
\\ 
\Gamma(\sigma \rightarrow \tilde{a} \tilde{a})
&=&\frac{\xi^2}{4\pi} \frac{m_\sigma m_{\tilde{a}}^2}{f_a^2}
\left(1-4\frac{m^2_{\tilde{a}}}{m_\sigma^2}\right)^{3/2},
\end{eqnarray}
where $m_{\tilde{a}}$ is axino mass.  In KSVZ model, the process
$\sigma \rightarrow aa$ overwhelms the other decay modes if $\xi \sim
O(1)$. We take $\xi=1$ unless otherwise noted. Then the total decay
rate is given by $\Gamma_\sigma\simeq \Gamma(\sigma \rightarrow a
a)$.\footnote{There exist the decay modes to gauge bosons. However,
  they are sub-dominant since they are suppressed by gauge coupling
  constant and the loop factor.}  On the other hand, in DFSZ model,
saxion interacts with Higgs doublets in $F$-term potential. Then
saxion can decay to the SM-like Higgs pair, whose partial decay width
is
\begin{eqnarray}
\Gamma(\sigma\rightarrow hh)=
\frac{k_\sigma}{4 \pi}\frac{\mu^4}{f_a^2m_{\sigma}}
\left(1-4\frac{m^2_h}{m_\sigma^2}\right)^{1/2},
\end{eqnarray}
where $k_\sigma$ is $O(1)$ constant, which is taken to be unity in the
later numerical evaluation, and $\mu$ is the $\mu$ parameter in the
MSSM superpotential.\footnote{Suppose there are two PQ fields,
  $\Phi_1$ and $\Phi_2$, and both of them get VEVs as $\langle
  \Phi_1\rangle=\langle \Phi_2\rangle$. If $\Phi_1$ couples to up- and
  down-type Higgses (denoted as $H_u$ and $H_d$, respectively) in
  superpotential, $\lambda \Phi_1 H_u H_d$ ($\lambda (\Phi_1^2/M_{\rm
    Pl})H_uH_d$), then $\mu=\lambda \langle \Phi_1\rangle$ ($\lambda
  \langle \Phi_1\rangle^2/M_{\rm Pl}$) and $k_{\sigma}=1$ ($2$). }
This decay mode dominates the total decay rate if $\mu\gtrsim
m_\sigma$. Saxion can also decays to sfermion pairs if kinematically
allowed. The decay rate for the process, however, is suppressed by
$\langle H_{u(d)}\rangle^2/\mu^2$ times Yukawa coupling constant
squared. ($\langle H_{u(d)}\rangle$ is the VEV of up (down)-type
Higgs.) Thus we ignore it.  For later calculation, we define the
branching fraction for axino pair production as
\begin{eqnarray}
{\rm Br}(\sigma \rightarrow \tilde{a}\tilde{a}) =
\frac{\Gamma(\sigma \rightarrow \tilde{a}\tilde{a}) }
{\Gamma_\sigma}. 
\label{eq:Br}
\end{eqnarray}
In KSVZ model, the branching ratio is simply given by ${\rm Br}(\sigma
\rightarrow \tilde{a}\tilde{a}) \simeq 8m_{\tilde{a}}^2/m_\sigma^2$ in
the limit $m_{\sigma}\gg m_{\tilde{a}}$. This is also true in DFSZ
model when $\mu\lesssim m_\sigma$.

\subsection{Axino production}

Axino is the fermionic component in the axion supermultiplet. Axino
can be produced in several ways; thermal production, non-thermal
saxion decay or the next-LSP (NLSP) decay. The production due to the
NLSP decay is negligible because the NLSP mainly decays to the SM
particles via RPV. The other two, {\it i.e.}, saxion decay and thermal
production, are potentially important.  In terms of yield variable
$Y_{\tilde{a}}\equiv n_{\tilde{a}}/s$ ($n_{\tilde{a}}$ is the number
density of axino), the resultant abundance of axino is expressed as,
\begin{eqnarray}
Y_{\tilde{a}}=Y_{\tilde{a}}^{\rm DEC}+Y_{\tilde{a}}^{\rm TH},
\end{eqnarray}
where $Y_{\tilde{a}}^{\rm DEC}$ and $Y_{\tilde{a}}^{\rm TH}$ are
contributions from saxion decay and thermal production, respectively.

$Y_{\tilde{a}}^{\rm DEC}$ is easily obtained.  Using
Eqs.~(\ref{eq:rho_sigma1}) and (\ref{eq:rho_sigma2}), the present
axino density due to saxion decay is given by
\begin{eqnarray}
Y_{\tilde{a}}^{\rm DEC}=
\frac{1}{4}d_X
\frac{{\rm max}\left\{T_{\sigma,{\rm osc}},\,T_{\rm dom}\right\}}{m_\sigma}
\left(\frac{\delta \sigma_{\rm ini}}{M_{\rm Pl}}\right)^2
{\rm Br}(\sigma\rightarrow \tilde{a}\tilde{a}).
\end{eqnarray}
Here we note that the produced axino is diluted due to the
late-decaying moduli, which is taken into account by the dilution
factor $d_X$. There is an entropy production due to saxion decay.
However, it is much smaller than the entropy production from
moduli. This is because saxion decays before moduli, the energy
density of saxion is smaller than that of moduli and that the
branching fraction to the MSSM-sector particles in saxion decay is
typically suppressed.  Therefore, the dilution due to saxion decay is
negligible compared to moduli decay.

The thermal production, on the other hand, is highly
model-dependent. It is described by Boltzmann equation,
\begin{eqnarray}
\dot{n}_{\tilde{a}} + 3Hn_{\tilde{a}}=C_{\rm prd}.
\end{eqnarray}
Here a dot means derivative with respect to the cosmic time and
$C_{\rm prd}$ is axino production rate per unit volume, which depends
on the axion model.  The solution of the Boltzmann equation in
radiation domination is given by (using $\dot{T}=-HT$)
\begin{eqnarray}
Y_{\tilde{a}}^{\rm TH}=\int dT \frac{C_{\rm prd}}{s(T)HT}.
\label{eq:Y_axino^THRD}
\end{eqnarray}
In KSVZ model, axino is mainly produced by thermal scattering or decay
of the particles in thermal plasma via strong interaction. For
example, the production rate due to scattering processes, such as
$\tilde{q} g \rightarrow \tilde{a}q$, $\tilde{g}g\rightarrow \tilde{a}
g$, is roughly estimated as $C_{\rm prd}\sim
\frac{\alpha_s^3}{f_a^2}n_{\rm MSSM}^2$ at high
temperature. ($\alpha_s$ is strong coupling constant and $n_{\rm
  MSSM}$ is the number density of the MSSM particle.)  Then from
Eq.~(\ref{eq:Y_axino^THRD}) the yield variable of axino is estimated
as
\begin{eqnarray}
Y_{\tilde{a},{\rm KSVZ}}^{\rm th} \sim 
O(10^{-3})\times \frac{\alpha_s^3M_{\rm Pl}T_R}{f_a^2}.
\label{eq:Y_KSVZ^TH_estimate}
\end{eqnarray}
It is seen that the axino production is the most active at the highest
temperature of the universe, {\it i.e.}, $T_R$.\footnote{Saxion decay
  reheats radiation during moduli domination. If the reheating
  temperature exceeds squark or gluino mass, then axino is also
  produced at the time of saxion decay.  The production is, however,
  negligible since the reheating temperature is much smaller than
  $T_R$ and that the production is suppressed by $T_{\rm dom}$ (see
  also later discussion).}  Thus the axino abundance is almost
determined by the production from thermal plasma before moduli
dominates the total energy, which guarantees that we have used
Eq.~(\ref{eq:Y_axino^THRD}).  More precise computation of the axino
production in radiation domination is done by Refs.~\cite{
  Covi:1999ty,Covi:2001nw,Choi:2011yf,Brandenburg:2004du,Strumia:2010aa}.
In our later numerical calculation we adopt the result given in
Ref.~\cite{Choi:2011yf} and fit their result as
\begin{eqnarray}
Y_{\tilde{a},{\rm KSVZ}}^{\rm th} \simeq 
{\rm min}\left\{ Y_{\tilde{a}}^{\rm eq},\ 
4\times 10^{-3} \alpha_s^3 \log(0.1/\alpha_s)
\left(\frac{T_R}{10^4~{\rm GeV}}\right)
\left(\frac{10^{11}~{\rm GeV}}{f_a}\right)^2\right\},
\label{eq:Yaxino_KSVZ}
\end{eqnarray}
where only QCD interaction is considered.\footnote{The fitting formula
  is applicable where $T_R\gtrsim 10^4\,{\rm GeV}$ and gluino or
  squarks are thermalized.}  It is seen that the estimate given in
Eq.~(\ref{eq:Y_KSVZ^TH_estimate}) roughly agrees with the expression.
$Y_{\tilde{a}}^{\rm eq}$ is the value when axion is thermalized, and
typically $Y_{\tilde{a}}^{\rm eq}\simeq 1.8 \times 10^{-3}$ using
$g_*=228.75$. The decoupling temperature $T_D$ can be estimated by
equating the scattering rate for the production process with the
Hubble parameter and it is obtained as
\begin{eqnarray}
T_D^{\rm KSVZ} \sim 10^8{\rm GeV}
\left(\frac{f_a}{10^{11}\,{\rm GeV}}\right)^2
\left(\frac{0.04}{\alpha_s}\right)^3,
\end{eqnarray}
which is consistent with Ref.~\cite{Rajagopal:1990yx}. Then axino is
thermalized when $T_R\gtrsim T_D^{\rm KSVZ}$. Since we assume $T_R\sim
T_{X,{\rm osc}}$, axino is thermalized in a wide range of the
parameter space.

Thermal production of axino in DFSZ model is different from the one in
KSVZ model.  As it is mentioned in
Refs.~\cite{Chun:2011zd,Bae:2011jb,Bae:2011iw,Choi:2011yf}, the
scattering process via strong interaction is suppressed at high
temperature. Instead, the production due to axino interaction with
Higgs and Higgsino or stop and top is effective.  For example, the
production rate for the processes, such as $\tilde{H}t\rightarrow
\tilde{a}t$, $\tilde{t}\bar{t}\rightarrow \tilde{a}h$, is roughly
$C_{\rm prd}\sim \frac{\mu^2}{\pi f_a^2T^2}n_{\rm MSSM}^2$. Thus axino
production occurs mainly in a lower temperature regime. From this fact
axino is produced after moduli dominates the total energy. The
solution given in Eq.~(\ref{eq:Y_axino^THRD}) can be used for the
axino production, except for using Eq.~(\ref{eq:H_XD}) for the Hubble
parameter.  As a result, the yield variable of axino during the epoch
of moduli domination is roughly obtained as
\begin{eqnarray}
Y_{\tilde{a},{\rm DFSZ}}^{\rm th}\Bigr|_{\rm XD} \sim
O(10^{-4})\times\frac{\mu^2}{f_a^2} \frac{M_{\rm Pl} }
{\sqrt{T_{\rm dom} \mu}}.
\label{eq:Y^TH_D_XD}
\end{eqnarray}
The contribution from decay gives the same order. Here we have assumed
that $T_R>\mu$.  It is seen that the resultant abundance is highly
suppressed by $T_{\rm dom}$. In addition, it is diluted by the late
moduli decay. Thus the contribution of the thermal production to axino
abundance before moduli decay is negligible in a wide parameter
range.\footnote{ Axino thermal production occurs after saxion decay if
  the decay reheats radiation to a temperature larger than $\mu$. The
  production results in the same order of yield variable given in
  Eq.~(\ref{eq:Y^TH_D_XD}). Therefore it is negligible for the same
  reason.}

Axino is also produced in the era of radiation domination after moduli
decay. Here $T_X$ plays the role of $T_R$ in the above discussion.  In
KSVZ model, however, the axino production is negligible because $T_X$
is smaller than gluino or squark mass in a wide parameter region, {\it
  i.e.}, processes, such as $gg\rightarrow\tilde{a} \tilde{g}$,
$qg\rightarrow \tilde{a}\tilde{q}$, are kinematically suppressed and
gluino and squarks are not thermalized.\footnote{This fact is crucial
  for moduli-induced baryogenesis. Otherwise produced baryon would be
  washed out. Axino can be produced via RPV interaction, such as
  $qq\rightarrow \tilde{a}q$. We have checked that this production is
  negligible in the parameter region that we are interested in.} On
the other hand, in DFSZ model, axino production may be substantial
since the production is effective at low temperature. The processes
without external stop, such as $\tilde{H}t \rightarrow \tilde{a}t$ or
$\tilde{H} \rightarrow \tilde{a}h$, contribute to the production
because the number density of stop is Boltzmann-suppressed.  Here we
assume $T_X > \mu$.  Then the production due to the scattering leads
to axino yield variable after moduli decay
\begin{eqnarray}
Y_{\tilde{a},{\rm DFSZ}}^{\rm th}\Bigr|_{\rm RD} \sim 
O(10^{-4})\times \frac{M_{\rm Pl} \mu}
{f_a^2}.
\end{eqnarray}
The contribution of Higgsino decay has the same order as one from the
scattering.  The above result roughly agrees with more accurate
numerical calculation in the literature.  Using the recent result
given in Ref.~\cite{Bae:2011iw}, the yield variable of axino is read
as\footnote{In Ref.~\cite{Bae:2011iw}, the result is given for the
  case relativistic stop is in thermal bath and its mass is larger
  than $\mu$. In such a case the yield variable is proportional to
  $\frac{M_{\rm Pl}\mu^2}{f_a^2 m_{\tilde{t}}}$.}
\begin{eqnarray}
Y_{\tilde{a},{\rm DFSZ}}^{\rm th} \simeq 
{\rm min}\left\{ Y_{\tilde{a}}^{\rm eq},\ 
 10^{-5}
\left(\frac{\mu}{1~{\rm TeV}}\right)
\left(\frac{10^{11}\,{\rm GeV}}{f_a}\right)^2\right\}.
\label{eq:Yaxino_DFSZ}
\end{eqnarray}
Here axino is thermalized when $T_D^{\rm DFSZ}\gtrsim \mu$, where
decoupling temperature $T_D^{\rm DFSZ}$ is given by
\begin{eqnarray}
T_D^{\rm DFSZ}\sim M_{\rm Pl}\frac{\mu^2}{f_a^2}.
\end{eqnarray}
Then the condition for axino thermalization becomes $\mu \gtrsim
10^4\,{\rm GeV} \left(f_a/10^{11}\,{\rm GeV}\right)^2$.  This
estimate is roughly consistent with Eq.~(\ref{eq:Yaxino_DFSZ}).

In summary, axino yield variable due to the thermal production is
given by
\begin{eqnarray}
Y_{\tilde{a}}^{\rm TH}= 
\left\{
\begin{array}{ll}
d_X  Y_{\tilde{a},{\rm KSVZ}}^{\rm th} 
& ({\rm KSVZ}) \\
Y_{\tilde{a},{\rm DFSZ}}^{\rm th} 
& ({\rm DFSZ}) 
\end{array} \right..
\end{eqnarray}
Then the density parameter of axino at present time is obtained by
\begin{eqnarray}
\Omega_{\tilde{a}}=m_{\tilde{a}} Y_{\tilde{a}}\, (\rho_c/s)_0^{-1}
= \Omega_{\tilde{a}}^{\rm DEC} + \Omega_{\tilde{a}}^{\rm TH},
\end{eqnarray}
where $(\rho_c/s)_0\simeq 3.6\,h^2 \times 10^{-9}\,{\rm GeV}$ for
$h\simeq 0.67$~\cite{Ade:2013zuv}.  Here we have split two
contributions for later convenience.  For example, in KSVZ model, they
are typically
\begin{eqnarray}
\Omega_{\tilde{a}}^{\rm DEC}h^2&\simeq&
0.43 \times \left(\frac{m_{\tilde{a}}}{20~{\rm GeV}}\right)^3
\left(\frac{10^{6}~{\rm GeV}}{m_{\sigma}}\right)^{3}
\left(\frac{m_{X}}{10^{10}~{\rm GeV}}\right)^{3/2}
\left(\frac{\delta \sigma_{\rm ini}}{M_{\rm Pl}}\right)^2, 
\label{eq:Omega_dec}
\\
\Omega_{\tilde{a}}^{\rm TH}h^2&\simeq&
0.084\times 
 \left(\frac{m_{\tilde{a}}}{20~{\rm GeV}}\right)
\left(\frac{m_{X}}{10^{10}~{\rm GeV}}\right)
\left(\frac{M_{\rm Pl}}{\delta X_{\rm ini}}\right)^2.
\label{eq:Omega_th}
\end{eqnarray}
Here we have used $T_{\rm dom}$ and $Y^{\rm eq}_{\tilde{a}}$ in the
estimation of $\Omega_{\tilde{a}}^{\rm DEC}$ and
$\Omega_{\tilde{a}}^{\rm TH}$, respectively. The expression of
$\Omega_{\tilde{a}}^{\rm DEC}$ can be applied in DFSZ model when
$\mu\lesssim m_\sigma$.

\subsection{Axion production}
  
We have seen that saxion mainly decays to axion pair. The produced
axion is relativistic thus behaves as radiation, which is so-called
dark radiation. The additional degree of freedom in radiation is
described in terms of the effective number of neutrinos $N_{{\rm
    eff}}=N_{\rm eff}^{\rm SM}+\Delta N_{\rm eff}$. Here $N_{\rm
  eff}^{\rm SM}=3.046$~\cite{Mangano:2005cc} is the prediction in the
standard model.  The result by Planck satellite~\cite{Ade:2013zuv},
combined with the measurements of the present Hubble parameter by
Hubble Space Telescope~\cite{Riess:2011yx}, gives $N_{\rm eff}=3.83\pm
0.54$ at 95\% C.L..  When the data from WMAP9~\cite{Hinshaw:2012aka},
the Atacama Cosmology Telescope~\cite{Sievers:2013ica} and the South
Pole Telescope~\cite{Story:2012wx,Hou:2012xq} are included, the
analysis gives $N_{\rm eff}=3.62^{+0.50}_{-0.48}$ at 95\%
C.L.\,\cite{Ade:2013zuv}. Though the current observations are
consistent with the SM value, the central values are slightly deviated
from the SM prediction.  We will see below that $\Delta N_{\rm eff}$
can be $O(1)$ in our scenario.

Referring to Refs.~\cite{Jeong:2012np,Choi:2012zna}, $\Delta N_{\rm
  eff}$ in our scenario is given by
\begin{eqnarray}
\Delta N_{\rm eff} = 3\, \left[\frac{\rho_a}{\rho_\nu}\right]_{\nu\,{\rm decp}}
=\frac{43}{7}\left(\frac{10.75}{g_*(T_X)}\right)^{1/3}
\left[\frac{\rho_{a}}{\rho_{R}}\right]_{X\,{\rm dec}}.
\end{eqnarray}
Here $\rho_\nu$ and $\rho_a$ are the energy density of neutrinos and
axion, respectively, and ``$\nu$ decp'' means the values at neutrino
decoupling. $\left[\rho_a/\rho_R\right]_{X\,{\rm dec}}$ is the ratio
of the energy density of axion produced by saxion and radiation at the
time of moduli decay. Using $\rho_\sigma\simeq \rho_a$ at the time of
saxion decay, it is straightforward to get
\begin{eqnarray}
\left[\frac{\rho_{a}}{\rho_{R}}\right]_{X\,{\rm dec}}
=\frac{4}{3}\,d_X
\frac{\left[\rho_\sigma/s\right]_{\rm osc}}{T_X}
\left(\frac{\Gamma_X}{\Gamma_\sigma}\right)^{2/3}
\end{eqnarray}
Assuming $\Gamma_\sigma\simeq\Gamma(\sigma\rightarrow aa)$ and using
Eq.~(\ref{eq:rho_sigma2}), $\Delta N_{\rm eff}$ is estimates as
\begin{eqnarray}
\Delta N_{\rm eff}\simeq 0.028
\left(\frac{10^{10}\,{\rm GeV}}{m_X}\right)^2
\left(\frac{m_\sigma}{10^{6}\,{\rm GeV}}\right)^2
\left(\frac{f_a/\xi}{10^{11}\,{\rm GeV}}\right)^{4/3}
\left(\frac{\delta \sigma_{\rm ini}}{M_{\rm Pl}}\right)^{2}.
\label{eq:DeltaNeffestimation}
\end{eqnarray}
In this paper we impose a conservative bound $\Delta N_{\rm
  eff}\lesssim 1$ on our scenario.

Axion is also produced by coherent oscillation when the Hubble
parameter becomes comparable to axion mass. If moduli decays before
the axion coherent oscillation, the abundance of the axion due to the
oscillation is the conventional value given in
Ref.~\cite{Turner:1985si}, {\it i.e.}, $\Omega_a^{\rm c.o.}h^2 \simeq
0.2 \theta_a^2 (f_a/10^{12}\,{\rm GeV})^{1.19}$. Here $\theta_a$ is
the initial misalignment angle of axion and it should be small in
order for axion not to overclose the universe if $f_a\gtrsim
10^{12}\,{\rm GeV}$.  The tuning of $\theta_a$ is possible when PQ
symmetry is broken during or before inflation.  Meanwhile, if PQ
symmetry is broken after inflation, the misalignment angle should be
replaced by $\pi/\sqrt{3}$. Then the tuning is impossible and $f_a$ is
severely constrained.\footnote{ In this case, the domain wall number
  should be unity. Even if $N_{\rm DW}=1$, axion is also produced from
  axionic string and axionic domain wall, which gives stringent
  constraint for the the decay constant, {\it i.e.}, $f_a\lesssim
  (2.0\mathchar`-3.8)\times 10^{10}\,{\rm
    GeV}$~\cite{Hiramatsu:2012gg}.  When the PQ symmetry is broken in
  during inflation, on the contrary, there is constraint from the
  isocurvature perturbation (see, {\it e.g.},
  Ref.~\cite{Hertzberg:2008wr}).  } In the case where moduli decay
after the coherent oscillation begins, the axion abundance is diluted
by moduli decay. Similar case is discussed in
Ref.~\cite{Kawasaki:2011ym}.  Since the axion abundance is
model-dependent and that we are interested in the axino DM scenario,
we simply assume that the axion energy density due to the coherent
oscillation is sub-dominant. It is straightforward to take into
account the axion abundance from the coherent oscillation and consider
mixed axion and axino DM.

\subsection{Axino stability}

In our model we consider RPV for moduli-induced baryogenesis. Through
RPV interaction, axino decays to SM particles even if it is the
LSP. The renormalizable RPV interaction in superpotential is
\begin{eqnarray}
W_{\slashed{R}_p}=\mu_i L_i H_u +
\lambda_{ijk}Q_iL_jD^c_k
+\lambda'_{ijk}L_iL_jE^c_k
+\lambda''_{ijk}U^c_iD^c_jD^c_k,
\end{eqnarray}
where $L_i$, $E^c_i$, $Q_i$, $U^c_i$, $D^c_i$ are chiral superfields
of left-handed lepton doublet, right-handed charged lepton,
left-handed quark doublet, right-handed up-type quark, right-handed
down-type quark, respectively. $i,\,j,\,k$ are generation indices. In
the present paper, we will take phenomenological approach to determine
the order of each RPV couplings as follows.  In our model baryon
asymmetry is generated through the RPV interaction.  Among the four
types of interactions, $U^cD^cD^c$ type is the most effective for
moduli-induced baryogenesis.\footnote{In Ref.~\cite{Cheung:2013hza} a
  simple baryogenesis is suggested in a minimal extension of the
  standard model by using $udd$ type higher dimension operator, which
  also contains a DM candidate. } For example, $\lambda''_{332}$ can
be order of unity evading from the severe constraint from proton
decay, and generate the observed baryon
asymmetry~\cite{Ishiwata:2013waa}.  The other couplings are partly
constrained phenomenologically (see, {\it e.g.},
Ref.~\cite{Barbier:2004ez}). Based on the facts, we simply consider a
case where at least one of $\lambda''_{ijk}$ is $O(1)$ and the others
are irrelevant.\footnote{Axino DM with different RPV operators is
  studied in, {\it e.g.},
  Refs.~\cite{Kim:2001sh,Hooper:2004qf,Chun:2006ss,Endo:2013si}.}

\begin{figure}[t]
  \begin{center}
    \includegraphics[scale=0.65]{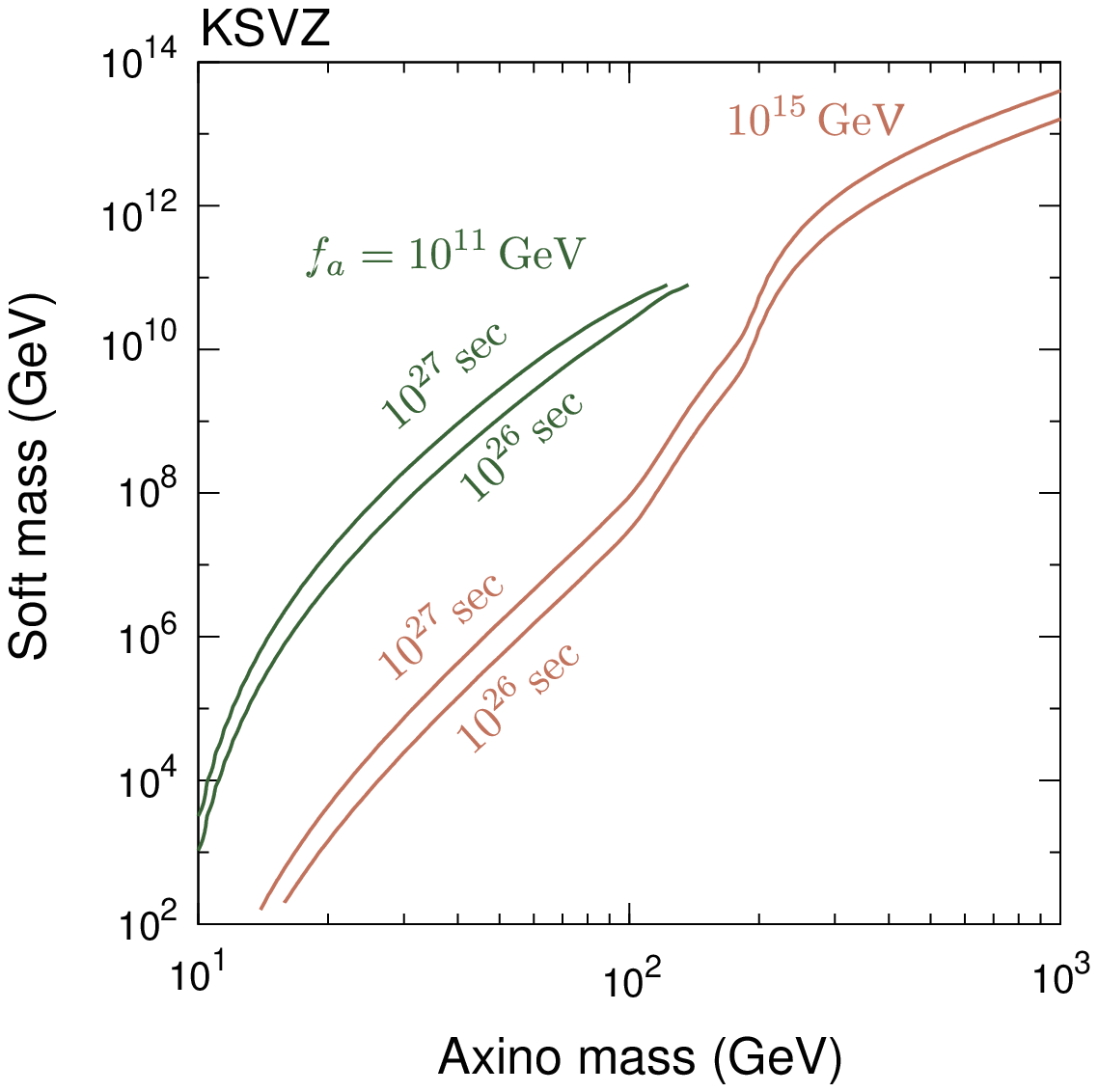}
    \includegraphics[scale=0.65]{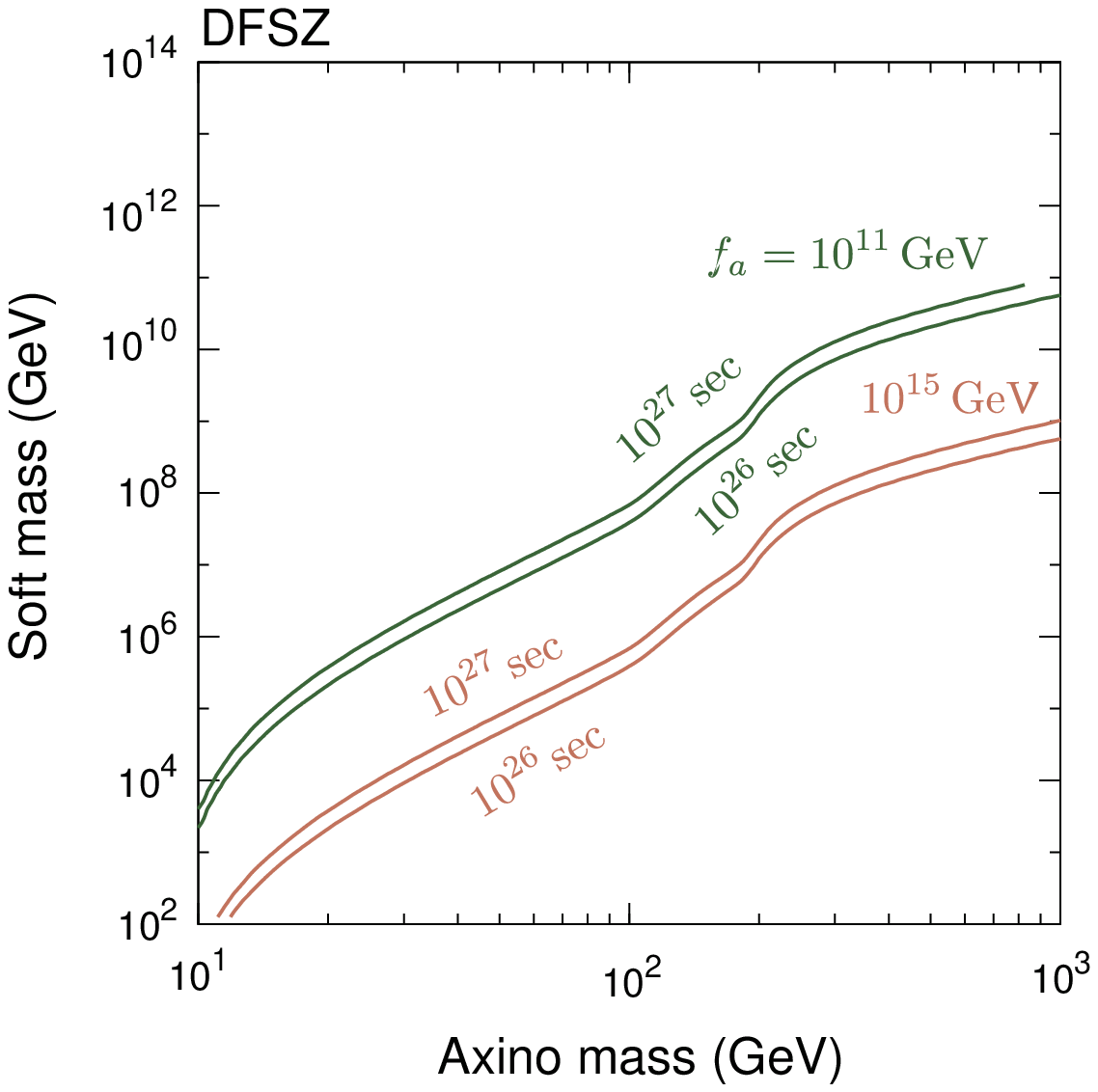}
  \end{center}
  \caption{\small Contour of axino lifetime. Left and right panels
    correspond to KSVZ and DFSZ model, respectively. The contours of
    $\tau_{\tilde{a}}=10^{26}$, $10^{27}\,{\rm sec}$ are depicted by
    taking $f_a=10^{11}$ (green), $10^{15}\,{\rm GeV}$ (red) in each
    panel. The plots are given in the region where the soft mass is
    less than $f_a$. Here we take $m_{\rm
      soft}=m_{\tilde{g}}=m_{\tilde{t}}$, $\lambda_{332}''=1$ and the
    others are zero.}
  \label{fig:tau}
\end{figure}

Axino lifetime is determined by the process $\tilde{a}\rightarrow
u_id_jd_k$. Relevant interaction for this process is 
dimension four axino-quark-squark coupling:
\begin{eqnarray}
{\cal L}_{\tilde{a}\mathchar`-q\mathchar`-\tilde{q}}=
g_{\rm eff}^{(L/R)}\tilde{q}_{L_i/R_i}\bar{q}_iP_{R/L}\gamma_5 \tilde{a},
\end{eqnarray}
where $P_{R/L}=(1\pm \gamma_5)/2$ and $\tilde{q}_{L_i/R_i}$ is
left-/right-handed squark.  Here quark and squarks are in the MSSM
sector.  In KSVZ model, although axino has no interaction with quark
and squark in the MSSM sector at tree level, the effective interaction
is induced at loop level. The effective coupling is given
by~\cite{Covi:2001nw}
\begin{eqnarray}
g_{\rm eff}^{(L/R)} \simeq 
\mp \frac{\alpha_s^2}{\sqrt{2}\pi^2}\frac{m_{\tilde{g}}}{f_a}
\log\left(\frac{f_a}{m_{\tilde{g}}}\right),
\end{eqnarray}
where $m_{\tilde{g}}$ is gluino mass.  On the other hand, in DFSZ
model, tree-level interaction exists in $F$-term potential, given
by~\cite{Chun:2011zd}
\begin{eqnarray}
g_{\rm eff}^{(L/R)} \simeq
\mp i \frac{m_q}{f_a}\left\{
\begin{array}{ll}
\cos^2 \beta & ({\rm for~up}\mathchar`-{\rm type~quark}) \\
\sin^2 \beta & ({\rm for~down}\mathchar`-{\rm type~quark}) 
\end{array} \right. .
\end{eqnarray}
Here $m_q$ is quark mass and $\tan \beta= \langle H_u\rangle/\langle
H_d \rangle$. In our model the soft SUSY breaking scale tends to be
large. To get the observed Higgs mass of around 126
GeV~\cite{Aad:2012tfa,Chatrchyan:2012ufa}, $\tan \beta \simeq 1$ is
required in the MSSM. Therefore we consider $\tan \beta = 1$, which
means that the axino decay is induced mainly by axino-top-stop
interaction.

In the following discussion we assume that all superparticle (except
Higgsino) in the MSSM sector have the same mass scale, which is
characterized by the soft mass $m_{\rm soft}$, {\it i.e.},
\begin{eqnarray}
m_{\rm soft} \sim m_{\tilde{f}},\ m_{\tilde{g}},\, {\rm etc.}
\end{eqnarray}
where $m_{\tilde{f}}$ represents sfermion mass. In the calculation of
axino lifetime, we use HELAS package~\cite{Murayama:1992gi}.

Fig.~\ref{fig:tau} shows contours of axino lifetime
$\tau_{\tilde{a}}$. Here we take $m_{\rm
  soft}=m_{\tilde{g}}=m_{\tilde{t}}$ ($m_{\tilde{t}}$ is stop mass),
$\lambda_{332}''=1$ and the other $\lambda''_{ijk}$ are
zero.\footnote{In the computation we ignored left-right mixing in
  squark sector for simplicity.  (Taking into account it is
  straightforward.) Here sbottom mediated diagram is neglected for
  simplicity by assuming stop is the lightest squark.} Via
$\lambda''_{332}$ axino decays to $tbs$ if kinematically allowed.  If
axino is lighter than top but heavier than $W$ boson, the final state
is $Wbbs$.  The final state becomes five body in which off-shell $W$
boson decays when $m_{\tilde{a}}\lesssim m_W$. (In the five-body final
state, we ignored fermion masses except for bottom quark.)  Those
behavior can be seen in the plot. When axino mass is around $W$ boson
mass and top mass, the lifetime is enhanced.  Then large soft mass is
required to suppress the lifetime.  In KSVZ model the lifetime is not
strongly suppressed by the soft mass compared to in DFSZ model. This
is due to a factor of gluino mass in the effective
$\tilde{a}$-$\tilde{q}$-$q$ coupling. Then the lifetime should be
suppressed by even larger soft mass.

\section{Results}
\label{sec:results}
\setcounter{equation}{0} 

Now we are ready to give numerical results.  Before showing the
results, we summarize the conditions which need to be satisfied for
our scenario:
\begin{eqnarray}
&{\it i)}&  T_X \gtrsim 10~{\rm MeV},
\\
&{\it ii)}& \tau_{\tilde{a}}\gtrsim 10^{26}\,{\rm sec},
\\
&{\it iii)}& \Gamma_{\sigma} > \Gamma_{X}.
\end{eqnarray}
{\it i)} and {\it ii)} 
are the phenomenological constraints, {\it i.e.}, moduli decays before
BBN and axino should not produce any exotic cosmic rays. The last one
is the condition in order for our scenario to work, {\it i.e.}, saxion
decays before moduli. In KSVZ model, it is simply given by
$m_X/m_{\sigma} \le 4.2\times 10^4 \left(\frac{10^{11}\,{\rm
      GeV}}{f_a/\xi}\right)^{2/3}$ in $m_{\sigma}\gg m_{\tilde{a}}$
limit.  It should be also reminded that we are interested in the mass
spectrum, such as
\begin{eqnarray}
m_{\tilde{a}} < (\mu,\, m_{\sigma},\, m_{\rm soft}) < m_{X}.
\label{eq:massspectra}
\end{eqnarray}

\begin{figure}[t]
  \begin{center}
    \includegraphics[scale=0.65]{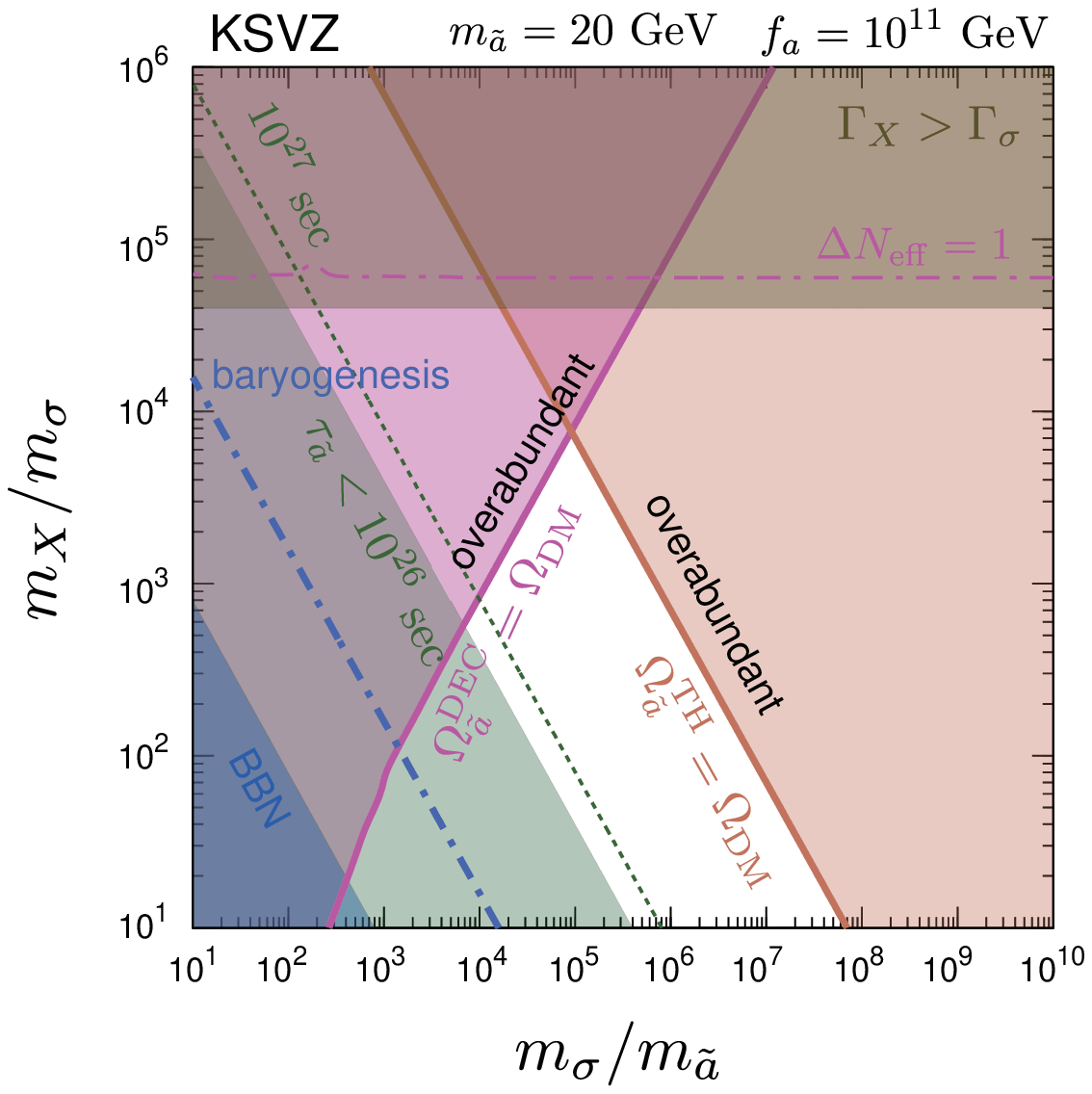}
    \includegraphics[scale=0.65]{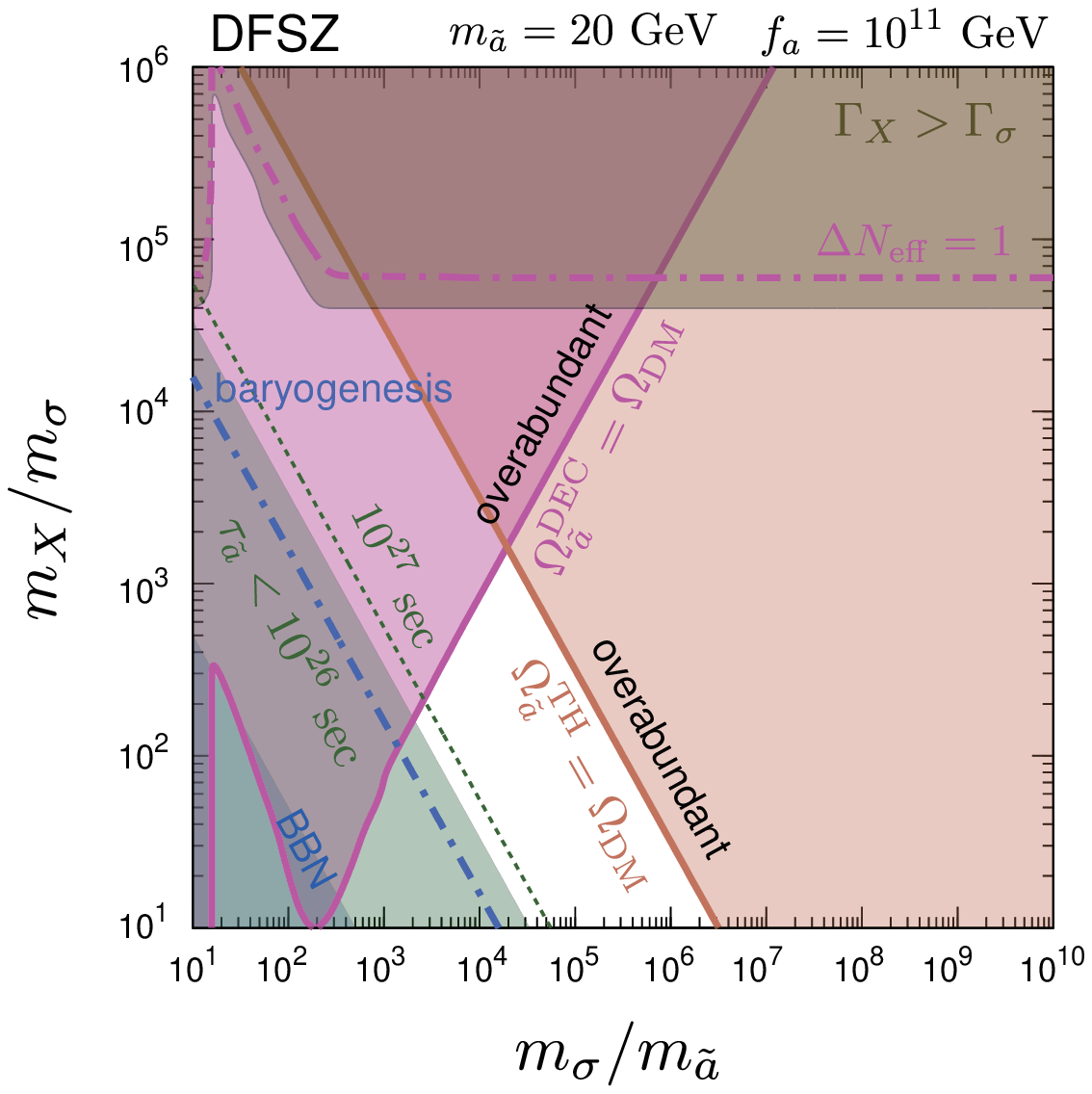}
  \end{center}
  \caption{\small Contours of $\Omega_{\tilde{a}}^{\rm
      DEC}=\Omega_{\rm DM}$ and $\Omega_{\tilde{a}}^{\rm
      TH}=\Omega_{\rm DM}$ on ($m_{\sigma}/m_{\tilde{a}}$,
    $m_{X}/m_{\sigma}$) plane.  Left (right) panel corresponds to KSVZ
    (DFSZ) model. We take $m_{\tilde{a}}=20~{\rm GeV}$,
    $f_a=10^{11}\,{\rm GeV}$, $\lambda''_{332}=(3\times 10^6\,{\rm
      GeV}/m_X)^{1/4}/2$ and $\delta X_{\rm ini}=\delta \sigma_{\rm
      ini}=M_{\rm Pl}$ for both models, and $\mu =10^2\,
    m_{\tilde{a}}$ for DFSZ model. In the plot shaded regions are
    excluded. ``BBN'' region is excluded due to $T_X< 10~{\rm MeV}$
    and ``overabundant'' means the region where
    $\Omega_{\tilde{a}}>\Omega_{\rm DM}$. The others are described in
    the figure. $m_X =3\times 10^6\,{\rm GeV}$ is drawn in (blue)
    dash-dotted line (also indicated as ``baryogenesis'') to show that
    successful baryogenesis is realized in region above the line.
    $\Delta N_{\rm eff}<1$ is satisfied in the region below the line
    $\Delta N_{\rm eff}=1$.  For reference, contour of
    $\tau_{\tilde{a}}=10^{27}\,{\rm sec}$ is also plotted in (green)
    dotted line.}
  \label{fig:contours}
\end{figure}

In Fig.~\ref{fig:contours}, contours of $\Omega_{\tilde{a}}^{\rm
  DEC}=\Omega_{\rm DM}$ and $\Omega_{\tilde{a}}^{\rm TH}=\Omega_{\rm
  DM}$ ($\Omega_{\rm DM}h^2=0.1196\pm 0.0031$ at $68\%$
C.L.~\cite{Ade:2013zuv}) are plotted on ($m_{\sigma}/m_{\tilde{a}}$,
$m_{X}/m_{\sigma}$) plane. Here we take $m_{\tilde{a}}=20~{\rm GeV}$,
$m_{\rm soft}=m_{X}/50$, $f_a=10^{11}~{\rm GeV}$, $\delta X_{\rm
  ini}=\delta \sigma_{\rm ini}=M_{\rm Pl}$. Left (right) panel shows
the result in KSVZ (DFSZ) model. In the plot of DFSZ model, we take
$\mu=10^2\, m_{\tilde{a}}$. For the determination of the axino
lifetime we take $\lambda''_{332}=(3\times 10^{6}\,{\rm
  GeV}/m_X)^{1/4}/2$ to explain the present baryon
density~\cite{Ishiwata:2013waa} and the others are taken to be zero.
$m_X\gtrsim 3\times 10^{6}\,{\rm GeV}$ should be satisfied for the
baryogenesis, which is also shown in dot-dashed line. In the figure
shaded regions are excluded.  Since axino mass is lighter than $W$
boson mass, the axino decay is five body.  We found that the
constraint $\tau_{\tilde{a}}\gtrsim 10^{26}\,{\rm sec}$ is much more
stringent than the BBN constraint, which excludes the lower mass
range.  The bound is stronger in KSVZ model due to the enhancement of
the effective $\tilde{a}$-$\tilde{q}$-$q$ coupling. However, it turns
out that the region $\Omega_{\tilde{a}}\simeq \Omega_{\rm DM}$ exists
in the valid parameter region for both models.  Two contributions,
$\Omega_{\tilde{a}}^{\rm DEC}$ and $\Omega_{\tilde{a}}^{\rm TH}$, have
different dependence on the mass parameters. In both models
$\Omega_{\tilde{a}}^{\rm DEC}$ is the same and well agree with
Eq.~(\ref{eq:Omega_dec}).\footnote{Except for low $m_{\sigma}$ range
  because $\sigma$ decay to Higgs pair changes ${\rm Br}(\sigma
  \rightarrow \tilde{a}\tilde{a})$.} As for the thermal production, on
the other hand, axino is thermalized in the region near
$\Omega_{\tilde{a}}^{\rm TH}\simeq \Omega_{\rm DM}$ but diluted
effectively to give the right amount in KSVZ model, which is
consistent with Eq.~(\ref{eq:Omega_th}).  In DFSZ model, axino is
copiously produced when $T_{X}$ become larger than $\mu$, which soon
becomes overabundant.  Therefore, the line $\Omega_{\tilde{a}}^{\rm
  TH}=\Omega_{\rm DM}$ locates near $T_X \sim \mu$. Regarding to axion
dark radiation, we have found that $\Delta N_{\rm eff}$ is less than
unity in the valid parameter region. To be concrete, the constraint
$\Delta N_{\rm eff}<1$ is always satisfied in the region
$\Gamma_\sigma>\Gamma_X$, independent of the parameters. (See
Eq.~(\ref{eq:DeltaNeffestimation}).)

We found the upper bound for axino mass. For large axino mass, the
bound from the lifetime becomes stringent. In order to suppress the
decay width of axino, large soft mass ({\it i.e.} moduli mass) is
needed. In KSVZ model, however, large axino mass and soft mass enhance
the thermal production of axino (see Eq.~(\ref{eq:Omega_th})).  Then
we found numerically
\begin{eqnarray}
m_{\tilde{a}}\lesssim 1\times 10^2\,{\rm GeV},
\label{eq:maxino_max}
\end{eqnarray}
by taking $f_a=10^{15}\,{\rm GeV}$. This bound can be also read from
Fig.~\ref{fig:tau}. When axino mass is larger than $O(100~{\rm GeV})$,
axino decays to $Wbbs$ where $W$ boson is on-shell, which leads to
enhance the decay rate. As a consequence, the constraint from the
lifetime and the overabundant bound destroy viable parameter
region. In DFSZ model, the bound from the lifetime becomes stringent
for large axino mass as well. In this case the thermal production
after moduli decay can be avoided if large $\mu$ is taken.  However,
the production during moduli domination is enhanced instead for large
$\mu$, which leads to overabundant axino. Then we found numerically
the upper bound for axino mass as
\begin{eqnarray}
m_{\tilde{a}} \lesssim 10^5\,{\rm GeV}
\end{eqnarray}
while taking $f_a=10^{15}\,{\rm GeV}$.

There is no lower bound for axino mass in this context. Then it is
possible to consider very large moduli mass. Let us suppose that
moduli mass is $O(10^{16}\,{\rm GeV})$. ($m_X\gtrsim 10^{16}\,{\rm
  GeV}$ is invalid since $T_X$ may be as large as the soft mass scale,
which may erase the baryon asymmetry.) With such a large $m_X$ and
small $m_{\tilde{a}}$, axino relic is mainly from thermal production.
Then axino with a mass of $O(10~{\rm keV})$ can be DM in KSVZ
model. In DFSZ model, axino DM should have a mass of
$m_{\tilde{a}}\sim O(0.1~{\rm MeV})$ when $f_a=10^{16}\,{\rm GeV}$ and
$\mu \sim 10^{14}~{\rm GeV}$, for example.  If the BICEP2 result is
confirmed, moduli mass should be larger than around $10^{16}\,{\rm
  GeV}$ in order for moduli to be stabilized.\footnote{See, {\it
    e.g.}, a recent work~\cite{Buchmuller:2014vda}, which takes into
  account the back-reaction effect.} Therefore our scenario is
compatible with high-scale inflation while stabilizing moduli.

Finally we discuss the experimental signatures involved in the
scenario. Near the region $\tau_{\tilde{a}}\sim
10^{26\mathchar`-27}\,{\rm sec}$, the decay of axino produces hadrons
and leptons, which may be observed as cosmic rays. Among them hadronic
decay products are especially constrained by cosmic-ray anti-proton
observation by PAMELA~\cite{Adriani:2010rc}.\footnote{See earlier
  works~, {\it e.g.}, \cite{Ibarra:2008qg,Ishiwata:2009vx}, which
  study cosmic-ray anti-proton from decaying DM.} When axino mass is
larger than order of a hundred GeV, a large amount of high energy
cosmic-ray anti-protons are generated. Then the cosmic rays will be
detected by AMS-02 experiment as an exotic signal, otherwise more
stringent constraint will be given. If axino mass is smaller, the
energy of the produced anti-proton gets smaller.  In such a low energy
range, the background cosmic ray increases. Thus it would be more
difficult to see the signal, depending on the lifetime. If axino is
lighter than 1 GeV, then axino becomes stable because it can not decay
to the SM fermions. However, proton decays to axino via the RPV
instead. As pointed out in Ref.~\cite{Dimopoulos:1987rk},
$\lambda''_{332}$ induces $ud\tilde{s}$ type coupling, $\kappa_{uds}$,
which is $O(10^{-7})\times \lambda''_{332}$. Then proton decay to
$K^+\tilde{a}$.  The decay rate of proton is estimated as
$\Gamma_{p\rightarrow K^+\tilde{a}}\sim
\frac{m_p}{16\pi}\left(\frac{\tilde{\Lambda}_{\rm QCD}}{m_{\rm
      soft}}\right)^4|\kappa_{uds}\,g_{\rm eff}^{(L/R)}|^2$. Here
$\tilde{\Lambda}_{\rm QCD}\sim 250\,{\rm MeV}$ is the QCD scale. Then
the lifetime of proton is estimated as $\tau_{p \rightarrow
  K^+\tilde{a}}\sim 3\times 10^{32}\,{\rm yr}
\left(\frac{f_a}{10^{10}\,{\rm GeV}}\right)^2
\left(\frac{m_X}{10^{10}\,{\rm GeV}}\right)^4
\bigl(\frac{4}{\log(f_a/m_X)}\bigr)^2 \bigl(\frac{250\,{\rm
    MeV}}{\tilde{\Lambda}_{\rm QCD}}\bigr)^4 $ in KSVZ model, and
$\tau_{p \rightarrow K^+\tilde{a}}\sim 5\times 10^{35}\,{\rm yr}
\left(\frac{f_a}{10^{10}\,{\rm GeV}}\right)^2
\left(\frac{m_X}{10^8\,{\rm GeV}}\right)^4 \bigl(\frac{250~{\rm
    MeV}}{\tilde{\Lambda}_{\rm QCD}}\bigr)^4 $ in DSVZ model. Here we
have used $m_{\rm soft}=m_X/50$ and $\lambda_{332}''\sim 0.07$ and $
0.2$ in KSVZ and DFSZ model, respectively. The current experimental
bound is $\tau_{p\rightarrow K^+ \nu}\ge 2.3\times 10^{33}\,{\rm
  yr}$~\cite{Kobayashi:2005pe}.  Therefore, proton decay experiment in
the future could be a test of this scenario even in high moduli mass
(soft mass) region.

In our scenario, lighter neutral Higgsino may be the LSP in the MSSM
sector and as light as $O(100~{\rm GeV}\mathchar`-1~{\rm TeV})$. If
the Higgsino is produced at a collider, it would decay inside the
detector via the $O(1)$ RPV. However, its decay width is suppressed by
the soft mass, thus it would decay from the interaction point.  Even
if the decay width of Higgsino is so suppressed by the soft mass that
the decay occurs far from the interaction point, the decay would be
observed. Then counting the number of the decay events, the lifetime
might be determined~\cite{Ishiwata:2008tp}. Then it may be possible to
probe the validity of this scenario in the high soft mass region.

\section{Conclusion}
\label{sec:conclusion}
\setcounter{equation}{0} 

In this paper we consider axino dark matter in large $R$-parity
violation.  While moduli dominate the universe after inflation, saxion
also oscillates coherently and eventually decays to produce large
amount of LSP axino. Axino is also produced thermally at the reheating
after inflation or lower temperature, depending on axion model.  Such
axinos are diluted by late moduli decay.  We have found that the axino
relic can give the correct amount to explain the present dark matter
abundance in both KSVZ and DFSZ models. Though axino is metastable due
to the large $R$-parity violation, its decay rate is suppressed the
axion decay constant, soft SUSY breaking mass or kinematics. Then the
lifetime can be longer in order axino not to produce exotic cosmic
rays.  With the large $R$-parity violation, baryon asymmetry is
generated in moduli-induced baryogenesis as well. Therefore the
scenario explains both dark matter and baryon existing in the present
universe.

\section*{Acknowledgment}
We are grateful to Wilfried Buchm\"{u}ller for discussions and giving
fruitful comments and suggestions during the research. We are also
thankful to Kwang Sik Jeong, Fuminobu Takahashi and Martin Winkler for
useful discussions.


\end{document}